\documentclass[journal]{IEEEtran}

\usepackage{color,soul}

\usepackage{ifpdf}

\ifCLASSOPTIONcompsoc
  \usepackage[nocompress]{cite}
\else
  \usepackage{cite}
\fi

\ifCLASSINFOpdf
  \usepackage[pdftex]{graphicx}
  \graphicspath{{../pdf/}{../jpeg/}}
  \DeclareGraphicsExtensions{.pdf,.jpeg,.png}
\else
  \usepackage[dvips]{graphicx}
  \DeclareGraphicsExtensions{.eps}
\fi

\usepackage{amsmath}
\usepackage{algorithmic}
\usepackage{array}
\ifCLASSOPTIONcompsoc
\usepackage[caption=false,font=footnotesize,labelfont=sf,textfont=sf]{subfig}
\else
\usepackage[caption=false,font=footnotesize]{subfig}
\fi
\usepackage{stfloats}
\usepackage{url}
\usepackage{color,soul}
\graphicspath{{../pdf/}{../jpeg/}}
\DeclareGraphicsExtensions{.pdf,.jpeg,.png}
\usepackage{mdwmath}
\usepackage{mdwtab}
\usepackage{eqparbox}
\newcolumntype{P}[1]{>{\centering\arraybackslash}p{#1}}
\usepackage[caption=false]{subfig}
\usepackage{booktabs}
\usepackage{amssymb,amsfonts}
\usepackage{mathtools}
\usepackage{multirow}
\usepackage{makecell}
\usepackage{algorithmic}
\usepackage[]{algorithm2e}
\usepackage{pifont}
\usepackage{verbatim}
\usepackage{tabularx}
\usepackage{textcomp}
\usepackage{float}
\usepackage[dvipsnames]{xcolor}
\usepackage{cite}
\usepackage[normalem]{ulem} 
\usepackage{orcidlink}

\begin{document}
\title{EdgeAIGuard: Agentic LLMs for Minor Protection in Digital Spaces}

\author{
Ghulam Mujtaba~\IEEEmembership{Senior Member,~IEEE}\,\orcidlink{0000-0001-9244-5346}, Sunder Ali Khowaja~\IEEEmembership{Senior Member,~IEEE}\,\orcidlink{0000-0002-4586-4131}, Kapal Dev~\IEEEmembership{Senior Member,~IEEE}\,\orcidlink{0000-0003-1262-8594}

\IEEEcompsocitemizethanks{ 
\IEEEcompsocthanksitem G. Mujtaba is with the Anderson College of Business and Computing, Regis University, Denver, CO, USA. E-mail: gmujtaba@ieee.org
\IEEEcompsocthanksitem S.A. Khowaja is with the School of Computing, Faculty of Engineering and Computing, Dublin City University, Ireland, and Faculty of Engineering and Technology, University of Sindh, Pakistan. E-mail: sunder.ali@ieee.org
\IEEEcompsocthanksitem K.Dev is with the Department of Computer Science, Munster Technological University, Cork, Ireland, and the Department of Institute of Intelligent Systems, University of Johannesburg, Auckland Park, 2006, South Africa E-mail: kapal.dev@ieee.org
}
}
\maketitle

\begin{abstract}
Social media has become integral to minors' daily lives and is used for various purposes, such as making friends, exploring shared interests, and engaging in educational activities. However, the increase in screen time has also led to heightened challenges, including cyberbullying, online grooming, and exploitations posed by malicious actors. Traditional content moderation techniques have proven ineffective against exploiters' evolving tactics. To address these growing challenges, we propose the EdgeAIGuard content moderation approach that is designed to protect minors from online grooming and various forms of digital exploitation. The proposed method comprises a multi-agent architecture deployed strategically at the network edge to enable rapid detection with low latency and prevent harmful content targeting minors. The experimental results show the proposed method is significantly more effective than the existing approaches.
\end{abstract}

 

\section{Introduction}\label{sec:intro}

Social media platforms have fundamentally transformed how individuals communicate, connect, and share information. It is not an exaggeration to say that social media has become integral to our daily lives. For minors, these platforms serve to form their identities, express themselves, and interact socially \cite{american2023health}. A recent study revealed that approximately 84\% of teenagers aged 13 to 17 actively use social media for an average of 4.8 hours daily \cite{trautman2024social}. Platforms like Snapchat, TikTok, and Instagram are easily accessible on devices like smartphones and wearables, allowing users to share their personal experiences while engaging with diverse content. Social media platforms have become crucial to minors' daily lives, enabling them to build virtual communities, make friends, explore shared interests, and participate in educational activities \cite{pewresearch2024}. That poses a significant risk and challenges for minors who use social media platforms for all those various activities.
\begin{figure}[!ht]
\centering
\includegraphics[keepaspectratio, width=\linewidth]{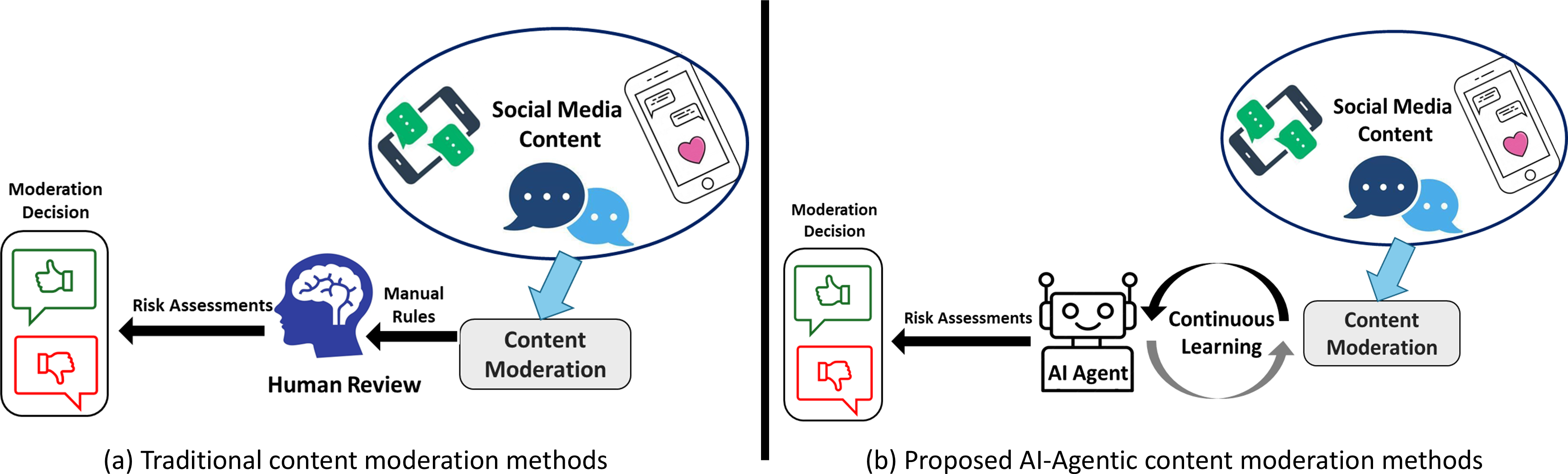}
\caption{\label{fig:fig1} Conceptual architecture of the (a) traditional and (b) proposed content moderation methods.}
\end{figure}
A recent report indicates that cyberbullying is quite common in social media spaces and affects approximately 59\% of minors \cite{wiederhold2024dark}. In the same study, it is also highlighted that 40\% of the affected minors report multiple incidents within a single year \cite{wiederhold2024dark}. It has also been revealed that online child/minor grooming cases have been on the rise for the past five years, i.e., an 82\% increase in cases \cite{canada_child_24}. Online child grooming refers to a relationship or friendship that is built online on social media specifically for manipulation, abuse, and exploitation \cite{trautman2024social}. Studies also reveal that 38\% of minors have at least faced some form of online harassment, and 45\% of minors have encountered explicit content unintentionally \cite{children2023lot}. Such incidents lead to severe psychological impact on minors, leading to anxiety, depression, and a decrease in academic performance \cite{american2023health}. Due to the rapid development of features on social media platforms, traditional content moderation methods are ineffective against the evolving tactics of exploiters. Therefore, advanced Artificial Intelligence (AI)--based solutions (refer to Figure \ref{fig:fig1}) are needed to tackle the suggested problem in an automated manner \cite{wenando2024optimizing, saleh2023detection}.

Recently, large language models (LLMs) have gained much attention from industries and the research community due to their ability to understand and analyze complex textual patterns, making them suitable for diverse applications \cite{khowaja2024chatgpt, mohamadi2023chatgpt}. Currently, a day hardly goes by without an announcement of LLM models that surpass their predecessors in terms of language understanding. Therefore, it is natural to consider that LLMs are best suited for detecting harmful online interactions and designing preventive measures for minors \cite{pan2024comparing, nirmal2024towards, najafi2024turkishbertweet}. The LLMs can identify subtle manipulation tactics, comprehend contextual nuances, and recognize evolving threat patterns that can be overlooked by traditional discriminative-based AI methods or rule-based systems. The ability of LLMs to process natural language and understand the patterns concerning contextual communication makes them a crucial technology for creating responsible, ethical, and safer digital spaces for minors. A safer and more accountable digital space can be realized using LLMs by designing an LLM-based moderation method that looks for sarcastic and bullying texts and identifies the patterns potentially used for online grooming. 

Researchers have tried to leverage the LLMs to realize online safety and detection of online grooming. However, the problems concerning edge implementation, latency, and the trade-offs between real-world applicability and detection performance remain large. For instance, Pahujani et al. \cite{BERT} proposed a Bidirectional Encoder Representations from Transformers (BERT)--based system that achieved 85\% accuracy for threat detection. However, the study did not meet real-time processing requirements. Wang et al. \cite{GPT3} proposed a Generative Pre-trained Transformers (GPT)--3-based implementation for grooming behavior identification. The study achieved 88\% accuracy on the identification task but faced privacy concerns as cloud-based processing was used to deploy the system. Wu et al. \cite{Transformers} proposed a transformer-based approach that achieved high accuracy, i.e., 90\%. However, it was far from real-world applicability due to the latency requirements. Some studies \cite{sen2024hatetinyllm} achieve a good trade-off, but only on provided datasets; hence, they fail to deal with the evolving nature of the threats as the method could not effectively handle context-dependent cases. The biggest issue for deploying LLMs for such applications is achieving the best trade-off while enabling its deployment on edge devices that could help realize the solution in a real-world context.

Agentic AI can address the aforementioned limitations by designing autonomous agents that can be deployed distributedly \cite{huang2024position}. Thus, it can be deployed at edge devices while incorporating self-optimization, adaptive intervention mechanisms, and contextual threat recognition. Unlike traditional AI methods that need fine-tuning, Agentic AI continuously refines detection strategies by updating decision-making parameters and constantly learning from emerging threats whenever the model needs to be updated. By leveraging Agentic AI with LLMs (especially with the distilled versions) \cite{thakkaragentmerge}, the detection models can be deployed at edge devices while improving response accuracy, enhancing proactive intervention, and mitigating the risks associated with online grooming and cyberbullying concerning minors. The multi-agent architecture also helps maintain user privacy through edge deployment, addressing the critical gaps associated with existing works.

In this paper, we present a new framework named EdgeAIGuard that utilizes Agentic AI with distilled versions of LLMs to safeguard minors from online grooming and various forms of digital exploitation. To be specific, the EdgeAIGuard utilizes three specialized agents, i.e., Sentinel, Context, and Intervention agents. The sentinel agent will be responsible for threat detection, the context agent will analyze historical data and the current Internet for contextual and semantic updates, and the intervention agent will be responsible for response generation. The proposed work is designed around edge-first architecture to ensure enhanced privacy and minimal latency. The specific contributions of the work are mentioned as follows:
\begin{itemize}
    \item A Novel multi-agent framework EdgeAIGuard for minors' online safety.
    \item An efficient strategy to deploy LLMs at the edge.
    \item An intervention system designed for evolving real-time threat prevention.
    \item Achieving state-of-the-art performance on edge devices. 
\end{itemize}

The remainder of the paper is organized as follows: Section \ref{sec:related_work} reviews the earlier related approaches to the proposed method. Section \ref{sec:methodology} outlines the methodology of the proposed EdgeAIGuard agent-AI content moderation method. Section \ref{sec:experiments} details the experimental setup, dataset information, and the experimental analysis. Finally, Section \ref{sec:conclusion} summarizes the proposed approach.

\section{Related Work}\label{sec:related_work}
Developing an automated technique to detect abuse, hate speech, and profanity in text presents a significant challenge. Numerous approaches have been proposed to tackle this issue, featuring various automated techniques \cite{jahan2023systematic}. A common method employs the Support Vector Machine (SVM) algorithm \cite{badjatiya2017deep, abozinadah2016improved} for identifying hate speech. Earlier research in \cite{badjatiya2017deep} compared machine learning and deep learning models against the SVM for hate speech detection. Another study focused on detecting hate speech in Arabic text from micro-blogging platforms such as Twitter \cite{abozinadah2016improved}. Traditional features like TF-IDF \cite{dinakar2012common, davidson2017automated}, bag of words \cite{pawar2018cyberbullying, ousidhoum2019multilingual}, and N-grams \cite{alakrot2018towards, malmasi2018challenges} have been extensively utilized in these approaches. For instance, \cite{dinakar2012common} developed user-interface models to identify cyberbullying through reflective features. In \cite{davidson2017automated}, a multi-class classifier was created to distinguish hate speech from offensive language using a crowd-sourced lexicon, with findings showing that racist and homophobic tweets were more often labeled as hate speech. In contrast, sexist tweets were typically seen as offensive. \cite{pawar2018cyberbullying} proposed a method for preemptively blocking cyberbullying messages. Furthermore, \cite{ousidhoum2019multilingual} introduced a multilingual approach for detecting hate speech across English, French, and Arabic, improving classification even with limited annotated data. \cite{alakrot2018towards} focused on offensive language detection in Arabic YouTube comments, emphasizing the significance of pre-processing techniques. Meanwhile, \cite{malmasi2018challenges} explored challenges in annotating hate speech and recommended better dataset quality through refined guidelines and increased annotator involvement.

Various types of word embeddings, including word-to-vector (word2Vec) \cite{kamble2018hate, faris2020hate}, Global Vectors for Word Representation (GloVe) \cite{rizos2019augment}, FastText \cite{wenando2024optimizing}, and embeddings from the language model (ELMo) \cite{zhou2020deep, saleh2023detection}, have been employed alongside architectures such as Convolutional Neural Network (CNN), Recurrent Neural Network (RNN), and Long Short-Term Memory (LSTM) to identify hate speech from textual data. These techniques are particularly relevant for social media platforms, encompassing tweets, news articles, posts, and messaging applications. For example, \cite{kamble2018hate} used English-Hindi code-mixed tweets with domain-specific embeddings. \cite{faris2020hate} applied word2Vec and AraVec in a hybrid CNN-LSTM model to classify Arabic tweets. \cite{rizos2019augment} investigated deep learning architectures, utilizing pre-trained embeddings like Word2Vec and GloVe while addressing class imbalance. \cite{wenando2024optimizing} focused on Indonesian social media, employing FastText with LSTM networks. Similarly, \cite{zhou2020deep} combined ELMo with CNNs to enhance classification through fusion strategies. Lastly, \cite{saleh2023detection} developed an unsupervised word embedding model and a BiLSTM-based classifier for improved detection accuracy.

The LLM-based approaches \cite{alatawi2021detecting, sen2024hatetinyllm, nirmal2024towards, de2024large, pan2024comparing, najafi2024turkishbertweet} surpassing CNN and RNN models in detecting hate speech content in textual data. Authors in \cite{alatawi2021detecting} explored the automatic detection of white supremacist hate speech on Twitter using the BERT architecture, which outperformed BiLSTM. Similarly, \cite{sen2024hatetinyllm} introduced HateTinyLLM for efficient hate speech detection, using fine-tuned, decoder-only tiny large language models (tinyLLMs) combined with Low-Rank Adaptation (LoRA) and adapter methods. In \cite{nirmal2024towards}, the authors presented SHIELD, an LLM-based technique that extracts rationales from input text to improve the interpretability of hate speech detection models.  In \cite{de2024large}, the researchers focused on enhancing hate speech detection with multilingual corpora and Cross-Lingual Learning (CLL) techniques, specifically for political discourse. Similarly, \cite{pan2024comparing} investigated LLM-based methods for detecting sexist and hateful online content through information retrieval, employing zero-shot and few-shot learning approaches. Finally, \cite{najafi2024turkishbertweet} introduced TurkishBERTweet, a LLM-based method trained on over 894 million Turkish tweets. Recent work in cross-technology authentication may inform secure agent communication and edge-level verification, such as AUTHFi \cite{wang2025authfi} and \cite{wang2024detection}.

Current research methods on hate speech detection using machine learning, deep learning, and LLM-based methods have significant limitations when deployed on edge devices. Traditional models such as SVM \cite{abozinadah2016improved} and feature-based methods \cite{ousidhoum2019multilingual, malmasi2018challenges} struggle with context-dependent hate speech and require extensive feature engineering. Deep learning models like CNNs, RNNs, and LSTMs \cite{rizos2019augment, faris2020hate, saleh2023detection} improve classification compared to traditional approaches but suffer from high computational costs and limited generalization beyond their training datasets. While LLM-based methods \cite{alatawi2021detecting, pan2024comparing, sen2024hatetinyllm} show promising results, they have significant latency issues, making real-time detection impractical. Due to high computational demands, existing LLM-based approaches are inefficient for edge deployment, raising concerns about privacy, cost, and real-time inference. Additionally, existing methods rely on specific datasets and fail to adapt to evolving online threats, limiting real-world applicability. We propose an Agentic AI-based approach designed on an edge-first architecture to address these challenges, ensuring enhanced privacy, minimal latency, and adaptive threat detection. The proposed EdgeAIGuard method is designed to detect grooming tactics such as trust building, isolation, desensitization, exclusivity, and manipulation via gifts or secrecy. It uses linguistic patterns and context history to identify such behaviors.
\section{Methodology}\label{sec:methodology}
\begin{figure}[t]
\centering
\includegraphics[keepaspectratio, width=\linewidth]{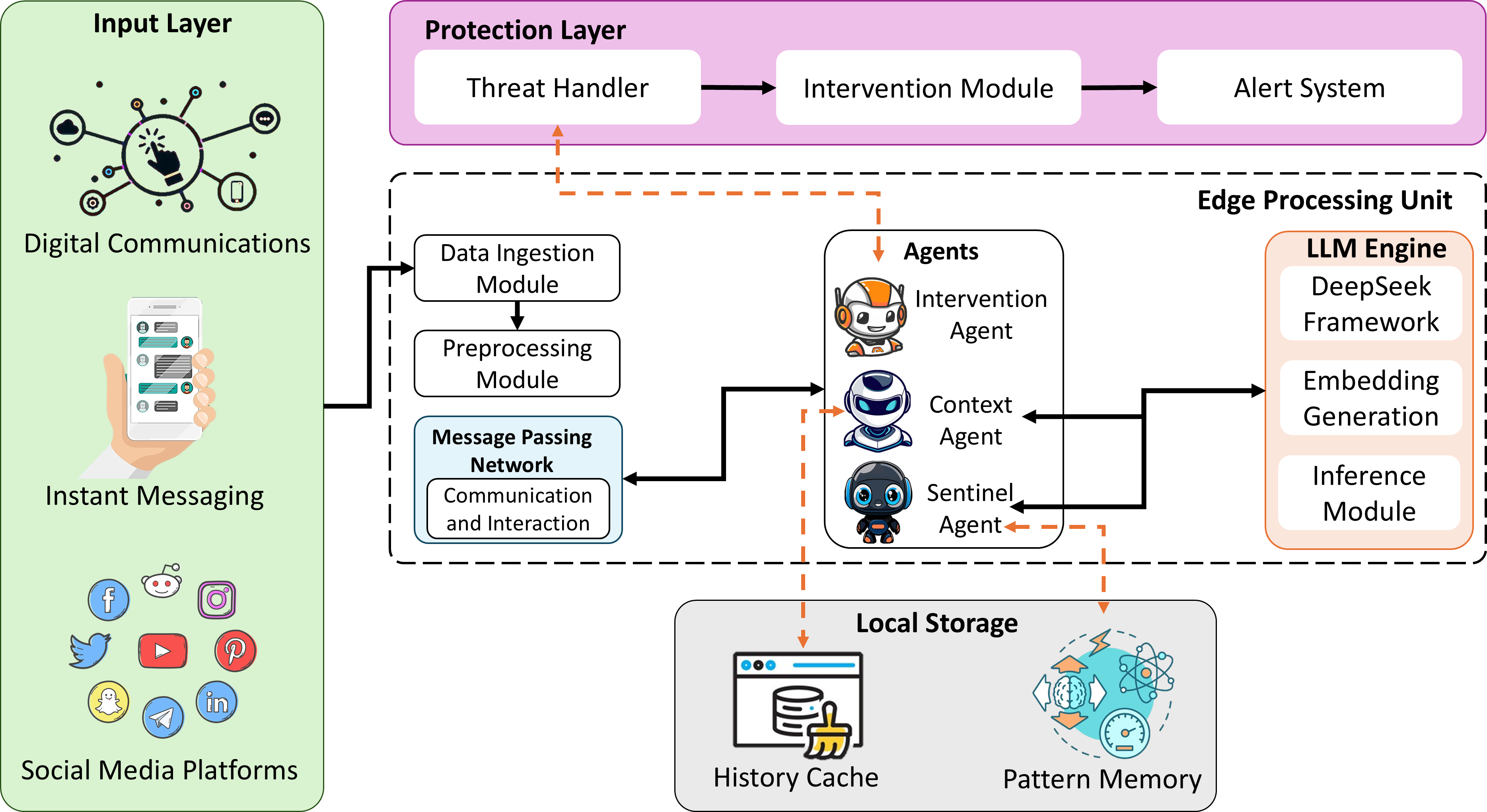}
\caption{\label{fig:fig2} Proposed EdgeAIGuard Architectural Workflow.}
\end{figure}

The proposed EdgeAIGuard's architectural workflow is shown in Figure \ref{fig:fig2}. The proposed framework is designed to protect minors using digital spaces through an innovative combination of edge computing and artificial intelligence agents. The framework comprises a multi-agent architecture deployed strategically at the network edge to enable rapid detection with low latency and prevent harmful content targeting minors.
The EdgeAIGuard framework comprises four essential layers, which are designed to optimize the protection pipeline for minors. The input layer collects data and processes information from various digital platforms. The form of information can be extended to the ones shown in Figure \ref{fig:fig2}, such as comments on a blog, discord chat, or any other platform that can be used for communication with minors. The edge processing unit forms the analytical core of the EdgeAIGuard framework, housing both the agent network and LLM engine. Edge deployment significantly reduces privacy risk by keeping user data on-device entirely. Only severe threats require external reporting. Local storage and processing eliminate the need for cross-border data transfer, enhancing GDPR and COPPA compliance. Context-aware decision-making is performed through local storage, which maintains emerging patterns and historical data for the AI agents. Finally, the intervention in the case of threat detection is executed by the protection layer in synchronization with one of the AI agents.

The proposed system adopts a sequential-parallel hybrid processing approach by allowing multiple specialized agents to process incoming data and information from the communications simultaneously. The hybrid approach maintains critical sequential dependencies essential for decision-making. It also optimizes the accuracy and processing speed, thus making the proposed EdgeAIGuard compliant with the rapid response time requirement for threat detection systems. EdgeAIGuard includes fallback logic to preserve core functionality in case of agent unavailability. If the Sentinel agent fails, the Context agent uses pattern memory and stricter thresholds to maintain threat detection. If the Context agent is missing, single-message assessments are weighted more heavily with a temporary history buffer. If the Intervention agent fails, predefined responses based on threat severity are used. This redundancy ensures that minor protection remains active even with partial system degradation.

The intervention agent is designed to execute protective actions tailored to factors such as the probability of harmful content, incident history, and user vulnerability. The intervention agent ensures interventions are proportionate and effective, balancing protection with minimal disruption. The following are some examples of interventions generated at different severity levels while using the intervention agent.

\subsection{Low Severity} When the system detects mildly inappropriate language, the intervention agent may display a safety message, such as, "The system detected some content that might not be appropriate. We advise being cautious and reporting anything that makes you uncomfortable." Such intervention aims to alert the user and educate without disrupting the conversation, while logging the interaction for further monitoring.

\subsection{Moderate Severity} If potential grooming tactics are identified, such as emotionally manipulating or attempting to isolate the minor, the intervention agent might pause the chat temporarily while sending a notification to the platform moderator or guardian stating, "We have detected behavior to isolate or manipulate the [minor's name]. The interaction needs reviewing from your end, accordingly." Such intervention will ensure the prompt involvement of guardians for additional oversight.

\subsection{High Severity} If a clear grooming attempt or explicit coercion is identified, the intervention agent will terminate the chat immediately and send an alert to the relevant authorities or the guardian with a message stating, "Potential exploitation or grooming attempt targeting [minor's name] detected. Your immediate attention is needed for the feedback. Details have been logged and encrypted for review." Such a response will ensure that user safety is prioritized while providing a rapid response to critical risk.

The aforementioned examples inferred from the agent describe its functionality, integration, and protection explicitly. Furthermore, the agent's ability to learn and adapt from feedback will enhance its accuracy in catering to evolving threats and users' preferences.

\subsection{Preliminaries and Problem Formulation}
We define the problem space formally by establishing a mathematical representation of digital communications. Let $C = \{c_{1}, c_{2}, c_{3}, ..., c_{M}\}$ represent the set of digital communications where each $c_{m}$ is a tuple as shown in Equation \ref{equ1}. 
\begin{equation}\label{equ1}
    c_{m} = (MC, SI, RI, TS, meta)
\end{equation}
Each set of digital communication in the EdgeAIGuard is treated as a multi-dimensional tuple encompassing the message content $(MC)$, sender information $(SI)$, receiver information $(RI)$, timestamp $(TS)$, and metadata, which provides contextual information as well. This rich digital communication representation helps the EdgeAIGuard capture both implicit contextual signals and explicit content to flag potential threats. We formulate the threat detection challenge in the EdgeAIGuard framework as a multi-output mapping function. The idea is not only to detect the threat but also to quantify the severity of the threat so that protective measures can be recommended accordingly. The formulation for threat detection is shown in Equation \ref{equ2}. 
\begin{equation}\label{equ2}
    CM \rightarrow \{0,1\} \times score_{sev} \times A
\end{equation}
where the $CM$ represents the class mapping, $score_{sev}$ refers to the threat severity score, and $A$ represent the set of recommended actions. The aforementioned threat detection formulation allows the EdgeAIGuard to generate nuanced responses that are compliant with detected threats' severity and specific nature. 

\subsection{Input Layer}
In order to capture the full spectrum of available information from digital communications, we implement a multi-modal feature extraction pipeline in the input layer. The input layer processes information from three parallel channels. Each channel provides specialized information from a different aspect of digital communication. The multimodal feature extraction pipeline is formulated as shown in Equation \ref{equ3}.
\begin{equation}\label{equ3}
    F(c_{m}) = \varphi(MC) \oplus \psi(SI, RI) \oplus \eta (meta)
\end{equation}
where the notations $\varphi, \psi,$ and $\eta$ refers to the text embedding function, user relationship embedding, and metadata embedding, respectively. The feature extraction process combines the analysis and assessment of the message content, the relationship between sender and receiver, and metadata evaluation. The text embedding function captures contextual and semantic information from the content. The relationship embedding analyzes the history and patterns of interactions between the parties communicating with each other. The metadata embedding extracts relevant information such as device information, timing patterns, and platform-specific features. The fusion of these features is performed using $\oplus$ operator, which preserves the relative importance of each information channel while creating a unified representation. A non-linear transformation is then used to capture complex interactions between different features. In this work, we have used the methods [ref for BERT and ROBERTA] to extract the embeddings, while the feature fusion is performed using [ref for Multihead attention]. We formally define the optimized communication embedding in Equation \ref{equ4}.
\begin{equation}\label{equ4}
    Emb(c_{m}) = \sigma(W_{c_{m}} F(c_{m}) + b_{c_{m}})
\end{equation}
where the notations $W_{c_{m}}$, $b_{c_{m}}$, and $\sigma$ correspond to the associated weight matrix, bias vector, and non-linear activation, respectively.

\subsection{Edge Processing Unit}
The core of the EdgeAIGuard is the edge processing unit that leverages the multi-agent system in conjunction with powerful LLM engine. The unit is designed to be operated on the network edge to provide rapid response time and enhanced privacy protection. We proposed three specialized agents for threat detection and prevention in the agent network. The first is the sentinel agent, which serves as the primary threat detector. The sentinel agent employs probabilistic modeling to identify potential threats in incoming communication. The formulation for the sentinel agent is shown in Equation \ref{equ5}. 
\begin{equation}\label{equ5}
    A_{sentinal} = argmax P(y|Emb) 
\end{equation}
In the Equation \ref{equ5}, $y$ represents the set of threat categories, while $P$ represents the probability distribution. Next, the context agent is designed to analyze the historical patterns and contextual information so that the network can better understand the communication dynamics. We define the context agent as shown in Equation \ref{equ6}.
\begin{equation}\label{equ6}
    A_{context}(Emb, H) = g(Emb \circ H )
\end{equation}
The notation $H$ represents the historical context vector, $g$ is the scoring function, and the operator $\circ$ represents the fusion operation. Lastly, we define the intervention agent that determines protective actions based on threat severity and contextual risks, using a weighted assessment of harmful content probability, incident history, and user vulnerability. Interventions may include safety messages, chat pauses, or alerts to guardians, tailored to the user's age and tone. The agent continually improves by evaluating response effectiveness and gathering feedback on false positives. We formally define the intervention agent in Equation \ref{equ7}. 
\begin{equation}\label{equ7}
    A_{intervene}(threat, score_{context}) = \pi(str | threat, score_{context})
\end{equation}
The notation $threat$ refers to the threat assessment, $score_{context}$ refers to the context score, $str$ refers to the intervention strategy, and $\pi$ represents the policy function through which the intervention agent will be optimized. \\
As described in Figure \ref{fig:fig2}, the inter-agent communication is performed through the message-passing network that enables information sharing among the agents to maintain system coherence. Each message between the agents carries weight information, which influences the collective decision-making process. We define the agents communication through message-passing networks formally in Equation \ref{equ8}.
\begin{equation}\label{equ8}
    MPN(A_{i}, A_{j}) = \rho(W_{mpn, ij} \cdot msg_{ij} + b_{mpn, ij})
\end{equation}
The above equation will consider the communication weight matrix $W_{mpn, ij}$, communication bias $b_{mpn, ij}$, and the message shared from $Agent_{i}$ to $Agent_{j}$. The resultant vector would be passed through message activation function $\rho$, accordingly. In this work, we employ DeepSeek framework as our LLM engine that processes communication while providing natural language understanding capabilities to the agent network. It should be noted that the EdgeAIGuard can use any LLM framework as its engine. However, DeepSeek is only used to realize this work. The LLM engine that processes communication is defined as shown in Equation \ref{equ9}.
\begin{equation}\label{equ9}
    LLM_{engine}(Emb) = DeepSeek(Emb, \theta)
\end{equation}
where $\theta$ represents the model parameters. 

\subsection{Local Storage}
The local storage system is implemented by leveraging dynamic memory architecture that maintains both long-term and short-term information necessary for performing threat detection. The dual-storage approach not only helps EdgeAIGuard to learn from past incidents but also helps the proposed system to adapt to emerging threats. We formalize the history cache and pattern memory in the Equation \ref{equ10} and Equation \ref{equ11}, respectively. 
\begin{equation}\label{equ10}
    HC(t) = \alpha \cdot HC(t-1) + (1 - \alpha) \cdot CS(t)
\end{equation}
The decay factor $\alpha$ is dynamic, defined as $\alpha = \text{base}_\alpha + (\text{sensitivity\_score} \times 0.1) - (\text{relevance\_score} \times 0.1)$, with $\text{base}_\alpha = 0.7$. This ensures that harmful patterns decay more slowly than routine content.
\begin{equation}\label{equ11}
    PM = {pm_{1}, pm_{2}, ..., pm_{Z}}
\end{equation}
The term $HC$ refers to the history cache, which utilizes a decay factor $\alpha$ and the current state represented by $CS(t)$, whereas the $pm_{z}$ refers to the set of patterns such that $(pattern, frequency, severity)$. The history cache maintains a record of recent system and communication states and is adjusted through time-weighting. The history cache also utilizes $\alpha$ to reduce the influence of older information gradually. However, it preserves important patterns accordingly. On the other hand, pattern memory stores recurring threat patterns, along with their severity and frequency. The memory enables the system to recognize evolving threat patterns and adapt responses. To protect privacy, all personal identifiers are anonymized using one-way hashing. We minimize and encrypt cached data using AES-256. Data processing is done entirely on-device, and all data is deleted after 30 days. The system complies with GDPR and COPPA by incorporating privacy-by-design principles, retaining data for only 7 days for minors under COPPA, and using features solely relevant to threats. The history cache retains anonymized interactions for 30 days, with older entries decaying gradually.

\subsection{Protection Layer}
Once a threat is detected, the protection layer becomes active for concrete protective actions. A hierarchical decision process for the protection layer considers multiple factors, including historical patterns, contextual risks, and threat severity. The hierarchical decision process is formulated in Equation \ref{equ12}. 
\begin{equation}\label{equ12}
\begin{aligned}
    TD(threat, score_{context}, risk) = \delta(W_{TD,1} \cdot threat \\
    + W_{TD,2}  \cdot score_{context} + W_{TD,3} \cdot risk)
\end{aligned}
\end{equation}
The notations $TD$ refers to the threat decision, $risk$ corresponds to the historical risks, and $\delta$ refers to the decision function. The weights for each of the terms are optimized so that the decision process for intervention is optimal. Subsequently, the intervention selection process is designed to balance the effectiveness of potential interventions while analyzing the impact on the user experience. To mitigate false positives, EdgeAIGuard employs context-aware thresholds that adjust based on communication history and detected patterns. The Sentinel agent uses ensemble scoring from content, metadata, and interaction history to reduce single-signal misclassifications. The intervention agent incorporates user feedback, and frequently flagged false positives are logged to suppress similar future alerts via dynamic memory adjustments. In this regard, the intervention process employs a utility-based approach, as shown in Equation \ref{equ13}. 
\begin{equation}\label{equ13}
    ISP = argmax[\zeta(a|CS(t))]
\end{equation}
The utility function is optimized considering the possible action $a \in A$ with respect to the current state $CS(t)$. The intervention process ensures that the protective measures are effective and proportionate to the detected threat and its severity.

\subsection{Objective Functions and Optimization}
The proposed EdgeAIGuard proposes a comprehensive multi-objective optimization approach in order to balance the competing goals and challenges. The composite loss function incorporates threat detection, context understanding, intervention effectiveness, and computational efficiency. The composite loss function is shown in Equation \ref{equ14}.
\begin{equation}\label{equ14}
\begin{aligned}
    loss_{composite} &= \lambda_{1} \cdot loss_{threat} + \lambda_{2} \cdot loss_{context} \\
    &\quad + \lambda_{3} \cdot loss_{ISP} + \lambda_{4} \cdot loss_{ce}
\end{aligned}
\end{equation}
where $loss_{ce}$ refers to the computational efficiency loss and $\lambda$s are the corresponding weights. The weights ($\lambda_1$–$\lambda_4$) were determined via grid search followed by adaptive tuning. We define each loss component in Equations \ref{equ15}, \ref{equ16}, \ref{equ17}, and \ref{equ18} as follows:
\begin{equation}\label{equ15}
    loss_{threat} = - \sum y log(\hat{y}) + R_{threat}
\end{equation}
\begin{equation}\label{equ16}
    loss_{context} = \|score_{context} - \hat{score}_{context}\|_{2} + R_{context}
\end{equation}
\begin{equation}\label{equ17}
    loss_{ISP} = - \mathbb{E}[str|\pi] + R_{ISP}
\end{equation}

\begin{equation}\label{equ18}
    loss_{ce} = \gamma(\tau + \omega) + R_{ce}
\end{equation}
The parameters $\gamma$, $\tau$, and $\omega$ refer to the function parameter, processing time, and resource usage. The $R$ terms correspond to the regularization terms associated with each loss. Each of the aforementioned loss functions addresses a specific aspect of EdgeAIGuard's system performance. The threat detection loss ensures the identification of harmful content to minors. The context loss helps the system understand complex communication patterns. The intervention loss ensures that the effectiveness of protective actions is optimized. The efficiency loss ensures the system complies with the edge network requirements and maintains real-time performance capabilities.

\subsection{Training Process}
The training of EdgeGuardAI architecture is carried out in an iterative manner, which gradually optimizes all the system components while maintaining the interdependencies of agents, digital communications, message passing network, LLM engine, and protection layer. This process combines supervised learning to detect threats and reinforcement learning to strategize interventions. We provide the training pseudocode in Algorithm \ref{algo1}. 

\begin{algorithm}[ht]
\label{algo1}
\caption{Training Process}
\KwData{Training Dataset $D$, Validation Dataset $V$, Maximum Epochs, Learning rate, Batch Size, Loss Weights, Regularization Coefficient}
\KwResult{Optimized Model Parameters $\theta = \{\theta_{sentinal}, \theta_{context}, \theta_{intervene}, \theta_{MPN}, \theta_{LLM}\}$}

\textbf{Initialization:}
\begin{itemize}
    \item Initialize agents and network parameters
    \item $\theta_{sentinal}, \theta_{context}, \theta_{intervene}, \theta_{MPN} \leftarrow \mathcal{N}(0, 0.01)$
    \item $\theta_{LLM} \leftarrow \text{DeepSeek-Lite-7B-Distilled}$
\end{itemize}

\For{each epoch $e$}{
    Set agents to training mode;
    Initialize epoch training loss: $epoch\_loss \leftarrow 0$;
    
    \For{each batch $b$ in $D$}{
        \For{each communication in batch $b$}{
            Extract features $F(b)$ (Eq. 3);
            Generate embeddings $Emb$ (Eq. 4); 
            $threat_{k}, prob_{k} \leftarrow A_{sentinal}(Emb(c_{k}))$;
            Retrieve history from cache (Eq. 10);
            $score_{context} \leftarrow A_{context}(Emb(c_{k}), H_{k})$;
            
            \For{each agent pair $(A_{i}, A_{j})$}{
                Update messages $MPN(A_{i}, A_{j})$;
            }
            $str_{k} \leftarrow A_{intervene}(threat_{k}, score_{context_{k}})$;
            $LLM_{output_{k}} \leftarrow \text{DeepSeek}(Emb(c_{k}), \theta_{LLM})$;
            $\tau_{k}  \leftarrow \text{processingtime}(c_{k})$;
            $\omega_{k} \leftarrow \text{resourceusage}(c_{k})$;
            Compute losses as shown in Eq. 14 - 18;
        }
        $batch\_loss \leftarrow \frac{1}{B} \sum loss_{k}$;
        $epoch\_loss \leftarrow epoch\_loss + batch\_loss$;
        $clip\_gradients(\theta, max\_norm = 1.0)$;
        
        \For{each communication in batch}{
            Update history cache with the current state (Eq. 10);
        }
        \If{new pattern detected}{
            $PM \leftarrow PM \cup \{(pattern, frequency, severity)\}$;
        }
        \Else{
            Update frequency and severity of existing patterns;
        }
    }
    $avgloss_{train} = \frac{epoch\_loss}{|D|}$;
}
\end{algorithm}

\section{Experimental Results}\label{sec:experiments}
In this section, we first present the experimental setup that outlines the tools used for conducting the experiments. We also provide the details regarding the dataset used in this study, ablation results, and comparative analysis with existing studies. This section aims to validate the system's effectiveness in threat detection and intervention when minors interact with digital environments. 

\subsection{Experimental Setup}
The EdgeAIGuard was evaluated using a combination of software simulations, controlled testbeds, and benchmark datasets to replicate real-world scenarios. We used NVIDIA Jetson AGX Orin to emulate the edge device, representing real-world deployment conditions due to its balance between computational power and energy efficiency. EdgeAIGuard runs on Jetson AGX Orin device with a minimum specification of a 6-core ARM CPU, 8 GB of RAM, 384-core GPU, 32 GB of storage, and 10W of power. We also used Intel Xeon Gold 6226R CPU paired with an NVIDIA A100 GPU, representing a high-performance cloud environment for baseline comparisons. As mentioned in the algorithm and architectural workflow, we would leverage DeepSeek-7B (distilled version) \cite{bi2024deepseek}, which is optimized for edge deployment without compromising performance. The DeepSeek-7B model was created using a teacher–student distillation framework. The 65B teacher model guided the student via output matching, intermediate representation alignment, and attention supervision. We employed INT8 quantization, attention pruning, and layer fusion. A distilled DeepSeek-7B model was used, achieving 87\% size reduction and 73\% latency reduction while preserving accuracy. For the model development and fine-tuning, we employed PyTorch \cite{pytorch_github} and HuggingFace Transformers \cite{Huggingface_transformers}.

The Agentic AI was implemented using Ray RLib \cite{RLlib}, which provides the implementation for efficient reinforcement learning and agent coordination. We leveraged TensorRT \cite{TensorRT} for model optimization to enhance inference speed and reduce latency on the edge device. We also adopted MQTT messaging protocol for lightweight and reliable agent communication. The EdgeAIGuard was tested within a mobile edge computing (MEC) emulator \cite{Unibo} to evaluate the performance under edge constraints. The traditional cloud-based solutions were conducted through parallel experiments on AWS EC2 instances. We used the batch size of 32 for training purposes, and the learning rate was set to 0.001 with a cosine annealing schedule. The ADAM optimizer was used with default parameters. We used 100 training epochs with early stopping. The loss weights, i.e. $\lambda$s were set to 0.4, 0.3, 0.2, and 0.1, respectively. The experiments were repeated five times with varying random seeds and the average results are reported in the results. The aforementioned setup ensures a thorough evaluation of EdgeAIGuard across varying environments for deployment and its applicability in real-world situations. 

\subsection{Dataset Details}
We mainly used two datasets to train the sentinel and context agents to identify implicit grooming attempts and detect harmful interactions. We utilized the online grooming dataset \cite{vogt2021early} that contains 1.2 million chat messages from known grooming incidents. These incidents were reported from real-world law enforcement case studies. The conversations in this dataset are labeled as predatory behaviors, coercion, and manipulative tactics. We use this dataset to train our context agent, accordingly. We also employ cyberbullying detection dataset \cite{ejaz2024towards, kaggle_dataset} which includes 4.5 million comments related to hate speech and cyberbullying. The dataset was collected from various social media platforms including Twitter, Reddit, and Discord. The dataset provides annotated comments for categories such as explicit content, threats, hate speech, profanity, and bullying. This dataset trains and assesses the sentinel agent's detection accuracy. Before model training, the aforementioned datasets are preprocessed using tokenization, normalization, and filtering techniques. For testing purposes, we utilized GPT-4 to generate real-time simulated threat interactions with annotations to test how EdgeAIGuard adapts to new and evolving threat patterns that were not part of the training data. All the experiments are carried out on the simulated threat interactions from GPT-4. To address dataset imbalance, we performed demographic audits and used GPT-4 to augment underrepresented groups. We applied stratified sampling across age, platform, and language categories. Adversarial training discouraged learning biased patterns.

\subsection{Experimental Analysis}
In order to evaluate the efficacy of the EdgeGuardAI, we use threat detection accuracy, false positive rate, and inference latency as our key performance indicators. The threat detection accuracy (TDA) will be measured to evaluate the correct classifications of harmful interactions in terms of \%. The false positive rate (FPR) will evaluate the capability of the agents to distinguish between harmful content and benign interactions. The inference latency will be measured in $ms$ to assess the real-time efficiency of the system. We first provide the threat detection performance using EdgeAIGuard in Table \ref{tab:tab1}, respectively. It can be noticed from the results that the EdgeAIGuard with the help of Agentic AI can perform better than the existing works, hence representing a 3.0\% improving the next best performing model. The results suggest that the EdgeAIGuard is more reliable in identifying threats with fewer false positives, which is a positive trait for real-world deployment.

\begin{table}[!ht]
\centering
\caption{Comparison of threat detection performance with existing works}
\label{tab:tab1}
\begin{tabular}{|c|c|c|c|c|}
\hline
\textbf{Model}       & \textbf{Precision} & \textbf{Recall} & \textbf{F1 score} & \textbf{Accuracy} \\ \hline
\textbf{BERT-based \cite{BERT}}       & 87.3 & 84.6 & 85.9 & 86.2 \\ \hline
\textbf{GPT-3 \cite{GPT3}} & 89.1 & 86.4 & 87.7 & 88.5 \\ \hline
\textbf{LLaMA \cite{LLAMA}}            & 90.2 & 87.8 & 89.0 & 89.6 \\ \hline
\textbf{HateLLM \cite{sen2024hatetinyllm}}          & 91.5 & 88.3 & 89.9 & 90.4 \\ \hline
\textbf{Transformer \cite{Transformers}}      & 89.7 & 87.1 & 88.4 & 89.1 \\ \hline
\textbf{EdgeAIGuard} & \textbf{93.7}      & \textbf{92.1}   & \textbf{92.9}     & \textbf{93.4}     \\ \hline
\end{tabular}
\end{table}

We then evaluate the performance of EdgeAIGuard in terms of real-time inference capabilities, which is crucial for such moderation and intervention systems, especially when deployed on edge devices with computational constraints. Currently, many AI models rely on cloud-based inference, leading to high latency and privacy concerns. We report the results for the real-time inference capabilities in Table \ref{tab:tab2}. The results also compare the existing baselines as well. The terms EL, CL, ET, and CT refer to edge latency, cloud latency, edge throughput, and cloud throughput, respectively. We cannot find edge-related results for GPT-3 as it is quite heavy. The results reveal the capability of EdgeAIGuard's real-world deployment on edge devices as it achieves the lowest latency on both edge and cloud environments, representing over 30\% improvement in edge latency from the next fastest model, i.e., HateLLM. The throughput of 10.2 requests per second on edge devices shows the suitability of EdgeAIGuard for real-time threat detection applications. 

\begin{table}[t]
\centering
\caption{Comparison of latency and throughput with the existing LLMs and methods}
\label{tab:tab2}
\begin{tabular}{|c|c|c|c|c|}
\hline
\textbf{Model}       & \textbf{EL (ms)} & \textbf{CL (ms)} & \textbf{ET (req/s)} & \textbf{CT (req/s)} \\ \hline
\textbf{BERT}       & 157 & 48  & 6.4 & 20.8 \\ \hline
\textbf{GPT-3} & N/A & 124 & N/A & 8.1  \\ \hline
\textbf{LLaMA}            & 213 & 67  & 4.7 & 14.9 \\ \hline
\textbf{HateLLM}          & 142 & 52  & 7.0 & 19.2 \\ \hline
\textbf{Transformer}      & 186 & 61  & 5.4 & 16.4 \\ \hline
\textbf{EdgeAIGuard} & \textbf{98}      & \textbf{42}      & \textbf{10.2}       & \textbf{23.8}       \\ \hline
\end{tabular}
\end{table}

Another important aspect of real-world deployment is the computational constraints the method faces. In this regard, we provide the analysis concerning the resource utilization on NVIDIA Jetson AGX Orin in order to provide the effectiveness of EdgeAIGuard in terms of memory usage (MU), GPU Utilization (GPU), Power Consumption (PC), and Model Size (MS) in comparison to existing works. The results for the resource utilization is provided in Table \ref{tab:tab3}. It should be noted that we did not use GPT-3 for this experiment due to its lack of support for the edge devices. It is clear from the results that the EdgeAIGuard achieves the lowest resource utilization across all metrics. For instance, EdgeAIGuard utilizes less memory, i.e. 13.3\% and lower power consumption, i.e. 6.9\% than HateLLM, which is the second-best performing method. It should also be noticed that the EdgeAIGuard maintains the best performance while having a significantly smaller model size, which is 287 MB. 

\begin{table}[!ht]
\centering
\caption{Comparison of resource utilization on NVIDIA Jetson AGX Orin with the existing LLMs and methods}
\label{tab:tab3}
\begin{tabular}{|c|c|c|c|c|}
\hline
\textbf{Model}             & \textbf{MU (MB)} & \textbf{GPU (\%)} & \textbf{PC (W)} & \textbf{MS (MB)} \\ \hline
\textbf{BERT}  & 1,258            & 78                & 12.4            & 438              \\ \hline
\textbf{LLaMA}       & 1,895            & 92                & 15.8            & 768              \\ \hline
\textbf{HateLLM}     & 962              & 72                & 10.2            & 325              \\ \hline
\textbf{Transformer} & 1,485            & 85                & 13.6            & 514              \\ \hline
\textbf{EdgeAIGuard}       & \textbf{834}     & \textbf{67}       & \textbf{9.8}    & \textbf{287}     \\ \hline
\end{tabular}
\end{table}

\begin{table}[!ht]
\centering
\caption{Ablation study comparing the performance of EdgeAIGuard with different agents. }
\label{tab:tab4}
\begin{tabular}{|c|c|c|c|}
\hline
\textbf{Configuration}          & \textbf{Accuracy} & \textbf{F1 score} & \textbf{Latency} \\ \hline
\textbf{(No Agent) LLM Only}    & 79.3 \%           & 78.5 \%           & 52 ms            \\ \hline
\textbf{Sentinel Agent only}    & 84.1 \%           & 83.2 \%           & 58 ms            \\ \hline
\textbf{w/o Intervention Agent} & 92.8 \%           & 92.1 \%           & 84 ms            \\ \hline
\textbf{w/o Context Agent}      & 88.7 \%           & 87.9 \%           & 82 ms            \\ \hline
\textbf{w/o Sentinel Agent}     & 85.2 \%           & 84.3 \%           & 76 ms            \\ \hline
\textbf{EdgeAIGuard}            & 93.4 \%           & 92.9 \%           & 98 ms            \\ \hline
\end{tabular}
\end{table}

We finally present the ablation study to show the contribution of each agent in the EdgeAIGuard framework. Table \ref{tab:tab4} reports the Ablation study, respectively. The results show that the sentinel agent provides the most significant impact to the EdgeAIGuard, leading to 8.2\% decrease if not used. The context agent also plays a crucial role in contributing to overall EdgeAIGuard's performance, however, it shows 4.7\% decrease which is less than that of the sentinel agent. While the intervention agent contributes marginally to the overall performance of the EdgeAIGuard, i.e., 0.6\%, it is essential for the appropriate response generation. 

EdgeAIGuard can support REST API by integrating existing moderation systems. It includes modules for platforms like Discord, Instagram, and TikTok to enable real-time alerts, webhook notifications, and family control extension. The proposed method is designed for scalable deployment across heterogeneous environments. The model architecture is modular and supports federated learning for decentralized updates. However, deployment at scale poses challenges such as variable hardware capabilities, update synchronization, and regulatory variations across jurisdictions. We mitigate these by providing device-specific model variants and centralized policy management interfaces for large platform partners. One of the limitations of the proposed work is the lack of evaluation on user-centered metrics such as adoption rates and user perceptions. As the evaluation of user-centered metrics is a large and diverse domain to cover, we would like to integrate the proposed work with a workable edge device in the future and provide it to users for real-world use. The aim would be to evaluate the user experience regarding trust, safety, and adoption. We intend to collect responses from minors and guardians on the system interventions. In addition, we would also like to track adoption rate, i.e., the number of users who enable the feature and retention over time, to understand the real-world impact.

\section{Conclusion}\label{sec:conclusion}
This paper proposed EdgeAIGuard, an innovative multi-agent framework for resource-constrained edge devices. The proposed approach is aimed at safeguarding minors from evolving online grooming and digital exploitation. The approach leverages a multi-agent architecture that includes Sentinel, Context, and Intervention agents. This facilitated comprehensive threat detection, contextual analysis, and the generation of appropriate responses while ensuring enhanced privacy and minimal latency. Experimental results from simulations, testbeds, and benchmark datasets showed that EdgeAIGuard significantly surpasses existing content moderation techniques. The edge-first architecture preserved contextual awareness through the local storage of emerging patterns and historical data. Thus, the proposed edge-based solution provided a scalable foundation capable of adapting to evolving exploitation tactics. This work advances the technical state of the art in content moderation and the broader objective of fostering safer online environments for minors. 

One of the limitations of the proposed work is the lack of evaluation of user-centered metrics such as adoption rates and user perceptions. As the evaluation of user-centered metrics is a large and diverse domain to cover, we would like to integrate the proposed work with a workable edge device in the future and provide it to users for real-world use. The aim would be to evaluate the user experience regarding trust, safety, and adoption. We intend to collect responses from minors and guardians on the system interventions. In addition, we would also like to track the adoption rate, i.e., the number of users who enable the feature and retention over time, to understand the real-world impact.

\bibliographystyle{IEEEtran}
\bibliography{IEEEabrv,manuscript}

\end{document}